\documentclass[11pt,a4paper]{article}
\usepackage[T1]{fontenc}
\usepackage[utf8]{inputenc}
\usepackage{lmodern}
\usepackage{microtype}
\usepackage{amsmath,amssymb}
\usepackage{graphicx}
\usepackage{wrapfig}
\usepackage{tikz}
\usetikzlibrary{arrows.meta,positioning,calc,shapes.geometric}
\usepackage{booktabs}
\usepackage[margin=2.55cm]{geometry}
\usepackage{caption}
\usepackage{xcolor}
\usepackage[colorlinks=true,allcolors=blue!50!black]{hyperref}
\usepackage[numbers,sort&compress]{natbib}
\usepackage{authblk}
\title{\bfseries After the Higgs: \\ Looking for the Universe Behind the Particle\\[0.3em]\large A perspective from Higgs Hunting 2026}
\author{Veronica Sanz}
\affil{Instituto de F\'isica Corpuscular (IFIC), Universitat de Val\`encia--CSIC, Val\`encia, Spain}
\date{November 2026}
\begin{document}
\maketitle
\begin{abstract}
Fourteen years after its discovery, the Higgs boson has changed roles. It was once the object of the search, but today it is an instrument with which we interrogate Nature's vacuum, test quantum field theory and look for new principles. This shift comes at a time when, unlike before the Higgs discovery, we have no theoretical argument guaranteeing new physics within experimental reach. At the same time, securing long-term support for fundamental science is becoming increasingly challenging. I revisit the path from a theoretically motivated discovery to a precision programme, explain why the Higgs is such an unusual quantum field, and argue that this same peculiarity makes it a gateway to cosmology and other sectors. I then turn to a less technical question: how should a field plan for exploration when no theorem guarantees the next discovery? My answer is to pursue precision without pretending to know in advance which measurement will matter, sustain a diverse experimental portfolio, protect theoretical risk-taking, and make the public case for fundamental exploration on its own terms.
\end{abstract}

\section{We know there is physics beyond the Standard Model}
A decade ago, the Higgs was the object of the search. Generations of scientists and engineers had worked relentlessly to build the largest particle collider ever constructed, capable of probing Nature at energies never before reached in a laboratory. And in 2012, after decades of searching, the LHC finally found it~\cite{ATLAS2012,CMS2012}.

Today, the LHC has produced more than 55 million Higgs bosons, and the particle that was once the object of the search has become an instrument for exploring the unknown. We still measure its couplings and test their agreement with the Standard Model, of course, but the questions driving these measurements have changed. We are no longer primarily trying to establish what this particle is. Instead, we are asking what it can tell us about the physics we have yet to understand. This transformation, from the object of discovery to a tool for exploration, is the starting point of this discussion.

Let me start with the obvious. The Standard Model is remarkably successful, and yet we know it is incomplete. Neutrinos have masses~\cite{SuperK1998}, most of the matter content of the Universe is of an unknown source, and  the observed asymmetry between matter and antimatter remains a mystery. The cosmological constant and cosmic acceleration confront us with a further problem whose resolution is far from clear. These are not anomalies produced by a difficult statistical combination. They are reasons to expect a more complete description of Nature. 

The status of inflation is different: it is an influential framework for early-Universe cosmology, not an empirically established new particle or interaction. And none of these questions disappears because Higgs measurements agree with Standard Model predictions.

What is less clear is whether the new physics we are looking for can be accessed with experiments we know how to build. And there is a second, logically distinct limitation that we should not forget: perhaps our theoretical language does not yet contain the right ideas. New physics might lie at energies beyond our experimental reach, manifest itself through exceedingly weak interactions, or require concepts fundamentally different from those we routinely employ.

These possibilities should make us modest, but they are no excuse for paralysis. If anything, uncertainty about what lies beyond our current paradigm should make us more, not less, daring. After all, paradigm shifts like the Higgs mechanism~\cite{EnglertBrout1964,Higgs1964,GHK1964} and gravitational waves~\cite{LIGO2016} were conceived decades before we developed the experimental capabilities to observe their consequences. In fundamental physics, ideas and the means to test them are often separated by generations. Think that when Brout, Englert and Higgs proposed the mechanism~\cite{EnglertBrout1964,Higgs1964} that would eventually become a cornerstone of the Standard Model, The Beatles had just arrived in America, and humanity had yet to set foot on the Moon. 

With these general comments, let us proceed with the discussion\footnote{The text follows a vision talk at Higgs Hunting 2026  V.~Sanz, ``After the Higgs: Looking for the Universe Behind the Particle,'' Higgs Hunting 2026 (17 September 2026), CERN video record \href{https://videos.cern.ch/record/3026972}{3026972}.}. I will start with a partly personal, and certainly selective, account of the years around the discovery. From there I will come back to the nature of the Higgs itself, before turning to a wider question: how do we organize a field around exploration when the next answer is not promised?

\section{When there was a beacon}
Those of us who worked on electroweak symmetry breaking before 2012 remember an unusual situation. We had something like a beacon. It did not tell us which road to take, and there was plenty of mud on the road, but at least it gave us a direction. The masses of the $W$ and $Z$, together with perturbative unitarity and the consistency of the quantum field theoretic description, told us that {\it something} had to happen around the TeV scale. A light Higgs was one resolution, new dynamics in longitudinal vector-boson scattering was another. The argument was not a mathematical guarantee that a particular resonance would appear in a particular detector. But it focused an experimental programme in a way that is rare in fundamental physics~\cite{LeeQuiggThacker1977}. 

Back then, I worked on alternatives without an elementary Higgs, including technicolour-like ideas~\cite{Weinberg1976}. Despite my enthusiasm for strongly coupled theories, inspired by Anderson's insights into superconductivity~\cite{Anderson1963}, I had painfully come to accept that the simplest interpretation of the electroweak data was an ordinary, relatively light Higgs boson. Occam's razor was not a proof, but it was hard to ignore, particularly when some of the alternatives were beginning to resemble creatures assembled in Dr Frankenstein's laboratory. 

And, yes, Nature chose the scalar. ATLAS and CMS discovered a new particle in 2012~\cite{ATLAS2012,CMS2012}, and its early behaviour was already strikingly close to that expected of the Standard Model Higgs. As my colleague John Ellis wittily put it, the new particle {\it walked and quacked like a Higgs}. 

The surprise since then is not only how much we have learned about the Higgs, but what the LHC itself has become. Once regarded, particularly by theorists, as a messy hadron collider primarily suited for discovery, it has been transformed by the experimental collaborations into an extraordinary precision instrument. The Higgs programme has evolved from establishing the particle's basic properties, such as its spin, parity and dominant decay modes, to increasingly sophisticated differential measurements, rare decay channels, couplings to second-generation fermions, and the search for Higgs pair production. Results remain consistent with the Standard Model within their measured uncertainties. Recent Run~2 combinations have reached few-percent precision on the inclusive Higgs signal strength, while individual coupling modifiers are generally less precisely determined~\cite{ATLASCMScouplings2016,ATLASNature2022,CMSNature2022,ATLASCombination2026,CMSCombination2026}. 

In other words, the story did not end with finding the particle. Discovery changed the task. The Higgs is no longer a candidate whose identity we are trying to establish. Instead, its properties are now being used to test the boundaries of the theory. The lack of a dramatic deviation should not conceal the extraordinary experimental achievement involved in making that possible.

But there is another, more fundamental reason to keep studying this particular particle. Even if every Higgs coupling were found to agree perfectly with the Standard Model, the Higgs would remain an extraordinary object. Establishing how closely it follows Standard Model predictions is one question. Understanding why Nature contains a light scalar in the first place is quite another.

\section{A quantum field theory unicorn}
\begin{wrapfigure}{r}{0.35\textwidth}
\centering
\vspace{-0.em}
\includegraphics[width=0.95\linewidth]{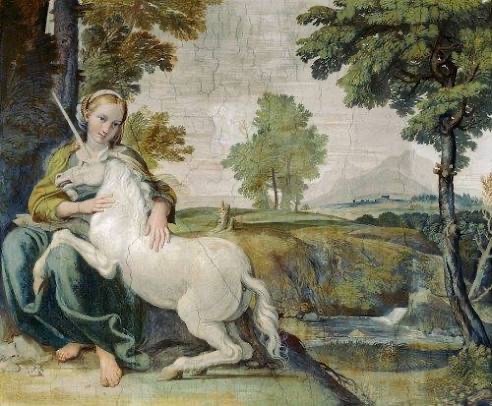}
\caption{Domenichino, \emph{The Maiden and the Unicorn}, c.~1604--05.}
\label{fig:unicorn}
\vspace{-0.5em}
\end{wrapfigure}

Why does the Higgs deserve so much attention? Not because it happens to be the last particle added to the Standard Model. It is unlike the other elementary particles we know. I used the unicorn in the talk (Fig.~\ref{fig:unicorn}) because it captures both the oddity and the temptation to domesticate it.

What makes the Higgs so unusual is not simply its spin or its role in electroweak symmetry breaking, but the way a fundamental scalar behaves in quantum field theory. In particular, while the masses of fermions and gauge bosons can be protected by symmetries, no analogous protection is automatically available for an elementary scalar.

For fermions, chiral symmetries can protect a small mass. Gauge symmetries similarly constrain vector-boson mass terms. An elementary scalar mass term, by contrast, is not forbidden by any of the Standard Model symmetries. As a consequence, the Higgs mass is not protected against quantum corrections from heavy particles, making it potentially sensitive to physics at energy scales far above its own. Indeed, in a theory containing additional heavy degrees of freedom with mass $M$ and couplings to the Higgs, threshold effects generically have the schematic form
\begin{equation}
\delta m_H^2\sim \frac{c}{16\pi^2} M^2 ,
\label{eq:threshold}
\end{equation}
up to coupling factors and model-dependent coefficients. This is a physical statement about sensitivity to heavy thresholds and it should not be confused with treating a quadratic regulator dependence as an observable. Nor does it establish that every possible ultraviolet completion must be tuned: symmetries, dynamics and particular ultraviolet structures can change the argument. The point is that the minimal theory does not itself explain why an elementary scalar should remain light in the presence of generic heavy physics~\cite{Giudice2008}.

This is what I mean by calling the Higgs a \emph{QFT unicorn}. We know a particle at approximately $125\,\mathrm{GeV}$ is there, and its observed properties are consistent with those of the Standard Model Higgs. Yet a fundamental scalar sits awkwardly with the intuition we have built from the quantum behaviour of fermions and gauge bosons. This puzzle invites us to look for new principles, but unfortunately it does not {\it guarantee} that the particles or phenomena associated with them will lie within our experimental reach.

How have theorists responded to this awkward situation? Much of beyond-the-Standard-Model phenomenology can be read as an attempt to turn that unicorn into a more ordinary horse. The broad idea is to associate the scalar with a structure whose quantum behaviour we have more reason to trust. Supersymmetry relates scalars to fermions~\cite{WessZumino1974}. Gauge-Higgs unification embeds Higgs degrees of freedom into higher-dimensional gauge structure~\cite{Hosotani1983}.  Composite Higgs scenarios use new strong dynamics and, often, a pseudo-Nambu--Goldstone interpretation~\cite{KaplanGeorgi1984}. These principles have many, many concrete realizations, typically with new partner states, resonances and modified Higgs interactions. Searches for such states and precision tests of the known Higgs are complementary, not competing, ways of asking the question of whether the unicorn has been domesticated.

Now, to be clear, naturalness is a theoretical problem, not {\it empirical} evidence for new phenomena. Yet attempts to its resolution provides important guiding principles to build extensions of the Standard Model. Indeed, the naturalness problem has motivated decades of theoretical developments and experimental searches, notably for Supersymmetry, many of them predicting new particles at accessible energies. And, so far, these particles have not been found, putting increasing pressure on the simplest models. But this does not mean that the original problem has disappeared, nor that Nature is obliged to follow our expectations of naturalness. The failure of particular predictions should make us reconsider our proposed solutions, not necessarily abandon the questions that motivated them.

\section{The Higgs as a Rosetta stone for the vacuum}
So far I have described the Higgs as a problem: a light scalar for which we seek an explanation. There is, however, another way of looking at precisely the same peculiarity. Because it is a scalar, the simplest object from a spacetime viewpoint, the Higgs can connect to sectors which would otherwise communicate only weakly with the familiar Standard Model particles. The feature that makes it sensitive to heavy physics also makes it potentially informative about that physics. And, unlike hypothetical new scalars, this is a field whose connection to electroweak symmetry breaking we already know is real and we are creating millions of it.\par
This is where I would change the metaphor. Instead of an awkward unicorn, think of the Higgs as a Rosetta stone: something familiar enough to study carefully, but capable of helping us interpret other, less accessible parts of Nature.

\begin{figure}[t]
\centering
\includegraphics[width=0.95\linewidth]{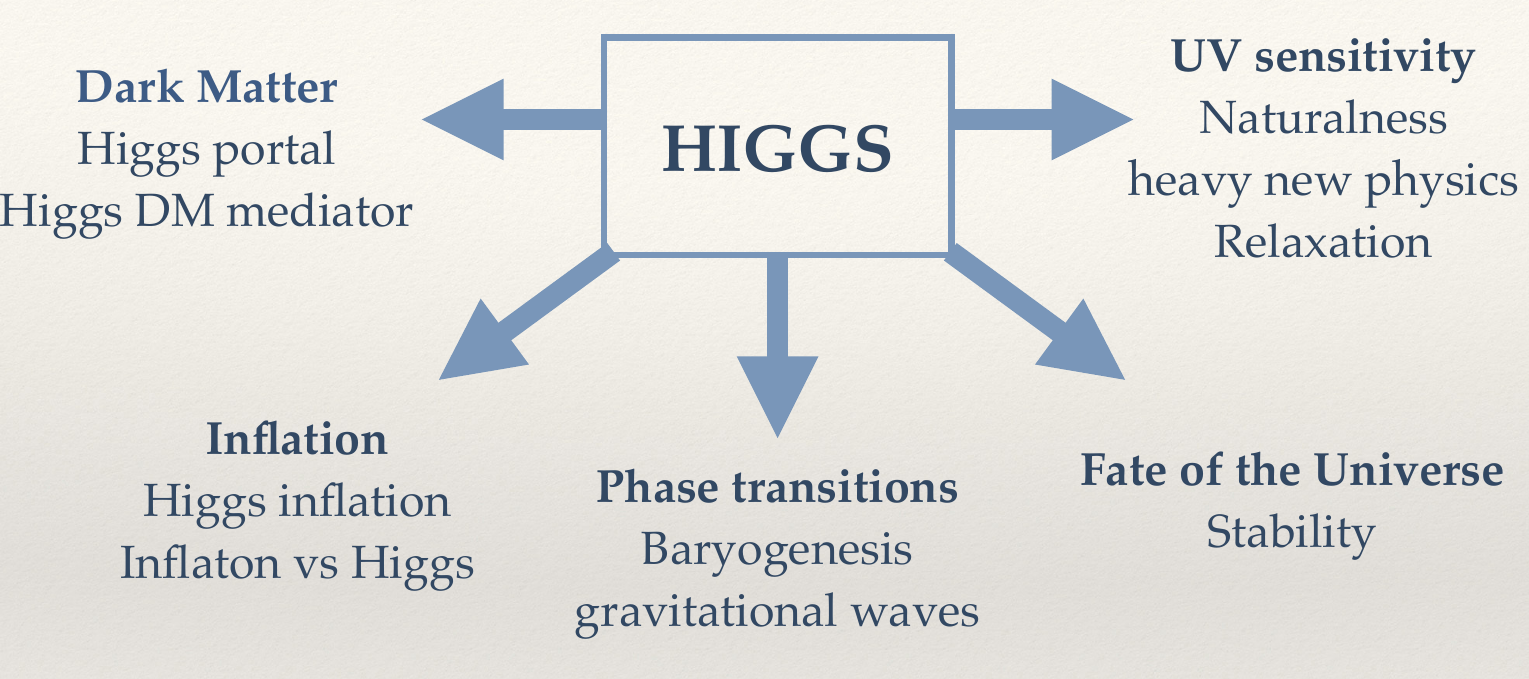}
\caption{The Higgs as a Rosetta stone: possible connections to dark matter, inflation, phase transitions, ultraviolet physics and the long-term fate of the electroweak vacuum. }
\label{fig:rosetta}
\end{figure}

Consider the possibilities sketched in Fig.~\ref{fig:rosetta}. A Higgs portal can mediate interactions with a dark sector~\cite{SilveiraZee1985}. Scalar dynamics may play a role in inflation, although the identification of the Higgs with an inflaton is model dependent and subject to important theoretical questions~\cite{BezrukovShaposhnikov2008}. The electroweak phase transition may connect Higgs-sector dynamics to baryogenesis and gravitational-wave signals if the physics goes beyond the minimal Standard Model~\cite{KuzminRubakovShaposhnikov1985,MorrisseyRamseyMusolf2012,Caprini2016}. Finally, extrapolating the Higgs potential to high scales raises questions about electroweak-vacuum stability and possible relations to early-Universe history~\cite{Degrassi2012}.

These are possibilities, linking  the Higgs with  big mysteries in Nature: dark matter, baryogenesis or cosmology. Their value is that they generate testable connections across different measurements. Invisible and exotic Higgs decays, differential distributions, Higgs pair production and effective-field-theory interpretations probe different aspects of this landscape. Cosmological and astrophysical observations constrain other parts. Neither side replaces the other. Their complementarity is precisely the opportunity.

The LHC has a privileged role because it produces the Higgs in a controlled environment and permits measurements whose systematic assumptions can be scrutinized. We should exploit that access as far as it will go, while recognizing that the phenomena to which the Higgs may connect need not live at collider-accessible energies. In this context, Effective Field Theory is one useful language for connecting precision observables to heavier dynamics, provided that its range of validity and the assumptions of a particular interpretation are kept explicit~\cite{Weinberg1979}.

Note that axions, axion-like particles and other scalar degrees of freedom offer similarly versatile connections. What distinguishes the Higgs is that its participation in electroweak symmetry breaking is established. For the Higgs,is not a generic portal imagined for convenience: it is a portal we already know exists in Nature.

That is the point of calling the Higgs a Rosetta stone for the vacuum. The claim is not that every outstanding question in cosmology reduces to Higgs physics. It is that the same well-characterized field enters several questions which we normally discuss in separate communities. The LHC may therefore tell us something important even when the most interesting consequence of a model appears far away from a collider.\par
There is still a long distance between recognizing this opportunity and organizing decades of measurements around it. And that brings me to the second part of the discussion. The question is no longer only what we should measure, but how we make room (scientifically and institutionally) for a programme whose most important outcome we cannot specify in advance.

\section{Fourteen years is not a long time}
We are making these decisions in a funding landscape that has become more demanding. There is a temptation, especially after a major discovery, to assume that the scientific success makes the case for the next large investment by itself. I do not think it does. We, as a scientific community,  have to explain what sustained exploration can deliver, while being candid about what it cannot guarantee. We all know that artificial intelligence, quantum technologies and other strategic priorities compete for public resources. And while it is perfectly reasonable to point out that these technologies owe much to fundamental science, and that hollowing out its foundations would be shortsighted, I believe that simply repeating that argument is not enough.

Funding is a social contract, not a theorem. No field is entitled to a fixed fraction of public resources because its questions are intellectually beautiful or because a previous generation delivered a spectacular discovery. I say this as someone who deeply loves fundamental physics. But if the terms of that social contract are changing, we should take the change seriously rather than arguing as though our scientific motivations automatically settle a political decision. The response cannot be merely to promise another discovery on a politically convenient timescale.

What sort of timescale is reasonable, then, for a precision programme whose questions evolve as it improves? One instructive, if imperfect, analogy is the cosmic microwave background. The signal detected by Penzias and Wilson~\cite{PenziasWilson1965} was not initially a precision map of the early Universe. Over roughly six decades, successive measurements of its spectrum, anisotropies and polarization turned it into one of the central probes of cosmology~\cite{Smoot1992,Planck2018VI}. The field did not stop once the signal was established. New questions became accessible as instruments, analysis methods and complementary observations improved.
\begin{figure}[t]
\centering
\includegraphics[width=\linewidth]{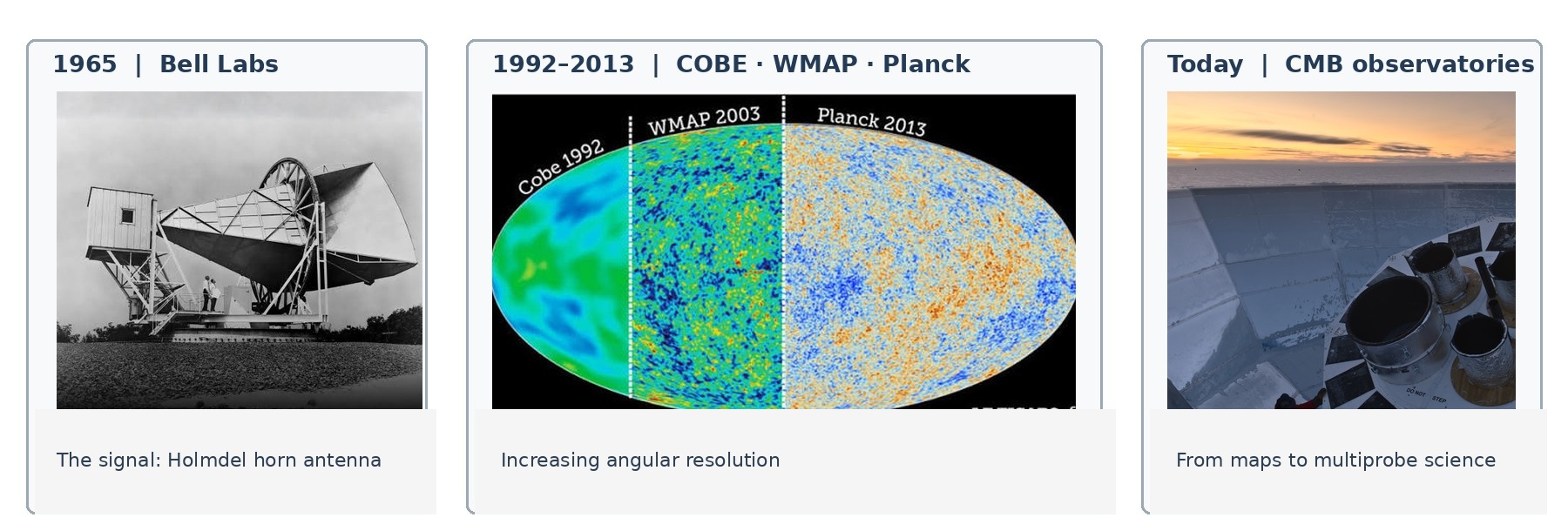}
\caption{A visual analogy across six decades: the Holmdel horn antenna at Bell Laboratories, the evolution from COBE to WMAP and Planck, and a modern CMB observing site. }
\label{fig:cmb}
\end{figure}

Today, the CMB is part of a much broader observational programme. Combined with measurements of galaxy clustering, baryon acoustic oscillations and gravitational lensing, it has allowed us to reconstruct the composition and evolution of the Universe with extraordinary precision, constrain the masses of neutrinos, and test the consistency of our cosmological model across very different epochs. More remarkably, by correlating CMB lensing maps with galaxy surveys, we can now trace how the large-scale structure of the Universe has grown over billions of years. What began as an unexplained excess of microwave noise has become a tool for mapping the evolution of the cosmos, from its infancy to the present day.

Of course, the CMB and the Higgs programmes are very different, and there is no reason to expect them to follow the same trajectory. But the comparison offers an important perspective: fourteen years after the Higgs discovery is not a particularly long time in the life of a major scientific programme. The history of the CMB reminds us that the transition from discovery to precision is not simply about adding decimal places. As measurements improve and new connections emerge, entirely new scientific questions become accessible, often ones that could hardly have been anticipated when the programme began.

The comparison is useful for another reason. A scientific programme lasting decades cannot survive intellectually on the questions that justified its first instrument. Long-lived infrastructures need evolving physics cases. We cannot assume that the interpretation imagined when an experiment was designed will still be the most valuable one a decade later. Experimental results should be documented and made available, wherever possible, in forms that allow them to be reinterpreted as new theoretical ideas emerge. We should avoid presenting measurements exclusively in terms of specific benchmark models, whose relevance may fade with time. In other fields, such as astronomy and space science, the public release of observational data is already standard practice, as illustrated by the extensive scientific archives maintained by NASA and ESA. Experimental collaborations already make considerable efforts in this direction, but preserving the flexibility to ask new questions of existing data should be regarded as an essential part of their scientific legacy.

\begin{wrapfigure}{r}{0.35\textwidth}
\centering
\vspace{-0.5em}
\includegraphics[width=0.98\linewidth]{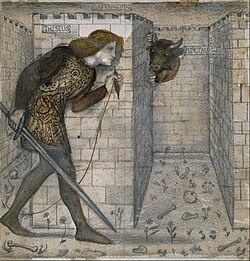}
\caption{Edward Burne-Jones, \emph{Theseus and the Minotaur in the Labyrinth} (1861).}
\label{fig:theseus}
\vspace{-1.8em}
\end{wrapfigure}

\section{Exploration without a no-lose theorem}
The labyrinth in Burne-Jones's image (Fig.~\ref{fig:theseus}) offers a useful counterpoint to the old beacon. This time, there is no clearly marked path towards the next discovery. We have to explore, take wrong turns, and occasionally find a way forward where we least expected it.

In contrast, we often describe the scientific method as a continuous dialogue between theory and experiment: theory proposes, experiment tests, and the results inform the next theoretical developments. This is a useful picture, and I regularly remind my students that physics is, above all, an \textit{empirical} science. But the history of physics is rarely as orderly as our diagrams suggest. Scientific progress does not always follow the script.

Experiments can reveal phenomena that were not anticipated by the dominant theoretical programmes, as happened with neutrino oscillations~\cite{SuperK1998}. The technologies developed to pursue scientific questions can also lead to entirely unexpected applications: the World Wide Web, born at CERN, is an obvious example. And sometimes theory advances by pursuing ideas with no foreseeable experimental test. Supersymmetry, for instance, emerged from attempts to understand how far the concept of symmetry could be extended in quantum field theory, long before it became the basis of a vast experimental programme.

The lesson is not that scientific progress is entirely unpredictable, but that we rarely know in advance which ideas, experiments or technologies will ultimately prove transformative.

These examples do not imply that priorities are impossible or that every proposal deserves equal support. They simply show the limits of planning discovery as if it were an engineering deliverable. In the absence of the electroweak beacon, we must be more explicit about what we can responsibly do now, even when we cannot name the eventual discovery.  I would draw a few practical consequences, separately for experiments and theory:

\paragraph{For experiments: precision is due diligence.}
Measure what is accessible as precisely as possible, even before we know why a particular measurement might matter. Precision is, in this sense, a matter of scientific due diligence. The LHC represents decades of extraordinary collective effort, and we have a responsibility to exploit its scientific potential to the fullest. This means reducing uncertainties, exploring what the machine can tell us, and ensuring that the results remain useful for questions we may not yet know how to ask. Precision is not, however, a substitute for a scientific question. Theory can widen the questions each result addresses, whether through effective field theory, unusual signatures or later reinterpretations that nobody anticipated when data were taken.

\paragraph{For long programmes: avoid guaranteed-success narratives.}
Colliders, detectors and other major instruments span many years. The physics case must be able to evolve. Reinterpret existing data, repurpose analyses, and be actively open to unexpected uses of detectors and techniques. Future experimental programmes should be justified by the scientific questions they can address and the measurements they can deliver, rather than by promises of discoveries we cannot guarantee. We can be confident in our experimental capabilities without pretending to know what Nature has in store for us. This uncertainty is inherent to fundamental exploration. Astronomers searching for exoplanets, for instance, can design increasingly sensitive instruments and develop sophisticated observational strategies, but they cannot guarantee the discovery of an Earth-like planet with signs of life. Nor should we be expected to guarantee the discovery of a new particle. What we can promise is to explore Nature with the best tools and ideas at our disposal, knowing that along the way we are likely to develop new technologies and methods whose applications may extend far beyond the scientific questions that originally motivated them.

\paragraph{For theory: disengagement is the danger.}
It is tempting to conclude, after years of null searches and impressively Standard-Model-like Higgs measurements, that the interesting ideas have already been tried. But experimentally, we have more ways of interrogating Nature than ever before. We have eyes in the sky, instruments buried deep underground, detectors at the bottom of the oceans and telescopes listening to gravitational waves. We collide particles at unprecedented energies, observe the Universe across the electromagnetic spectrum, and probe phenomena spanning an extraordinary range of scales, from the subatomic to the cosmological. As theorists, we are living in a privileged era, with an unprecedented wealth of observations to challenge our ideas and inspire new ones. Theorists should help shape those questions and be ready to interpret a surprise. This requires close engagement with measurements, while leaving room for approaches that do not arise from immediately fitting the next exclusion plot.

\paragraph{For younger researchers: protect intellectual risk.}
Original ideas need difficult questions and time in which to fail. Senior researchers can afford to investigate improbable directions in ways that early-career scientists often cannot. Younger colleagues face considerable pressure to publish regularly, secure funding and build a competitive CV, often while moving between short-term contracts. Under these conditions, pursuing an unconventional idea with uncertain prospects can become a serious career risk, even when the scientific question is worth exploring. As senior researchers and supervisors, we have a responsibility to create room for intellectual risk-taking, so that the pressure to establish a career does not come at the expense of originality. If every project must promise a publication on the timescale of the next contract, then we should not be surprised if the field becomes conservative. And this is precisely what we can least afford at a time when the direction of the field is uncertain and progress may depend on questioning established assumptions and exploring genuinely new ideas. Protecting a modest amount of time and resources for work that sounds unusual is part of sustaining a creative theory community. Note that this is not a recommendation to stop demanding technical rigor, just a reminder  that rigor and intellectual risk are compatible, a combination  very much needed now.

\section{The case for exploration}
All these suggestions come back to one uncomfortable fact. We have no no-lose theorem today. Nothing requires a new particle to appear within our lifetimes. In retrospect, the comparatively clear electroweak target before the Higgs discovery was itself unusual. The history of fundamental research is not a sequence of guaranteed discoveries with known delivery dates.

This should change how we describe what we do. Particle physics is a worthy human endeavour without a promise to discover particle $X$ by year $Y$. This uncertainty is hardly unique to particle physics. Many areas of fundamental research pursue questions whose answers may remain beyond our reach for generations. Neuroscience aims to understand the workings of the human brain, while astrobiology searches for life beyond Earth. Neither field can guarantee that its ultimate goals will be achieved, yet this does not diminish the value of the questions or prevent the development of ambitious, long-term research programmes.  Particle physics should be capable of making a similarly candid case for the importance of its questions, its methods and its potential to surprise us.

My preference for a long-term strategy is diversification: push precision within large collaborations, including collider and flavour programmes, while supporting a broad portfolio of smaller, focused experiments. This is not because every approach is equally promising. It is because we do not know which observable, energy range or experimental principle will provide the next clue. In my opinion, concentrating all resources in one interpretation of the future is a large scientific gamble.

That diversified strategy is a scientific answer to uncertainty. But it is not, by itself, an answer to the people who are asked to support it. We do not work in an ivory tower, detached from the concerns and priorities of the society that funds our research. Yet, at times, we may give precisely that impression, particularly when we discuss the future of our field as though its importance and entitlement to public resources were self-evident.

This brings us back to the social contract. We need to devote real effort to explaining why fundamental science deserves sustained public support, without promising discoveries on demand or technological returns on an unrealistic timescale. We have a compelling case to make: the value of exploring Nature, of asking difficult questions and building the instruments to address them, of developing knowledge and skills, and of inspiring new generations to participate in this intellectual adventure. But we cannot expect that case to make itself. We should work {\it as hard} to earn public support as we do to design the next experiment. The Higgs took nearly half a century to move from a theoretical idea to an experimental discovery. We do not know what the next half-century will bring, but I hope we will give future generations the freedom, the tools and the ambition to find out.

\end{document}